\documentclass[a4paper]{jpconf}
\usepackage{graphicx}
\begin{document}
\title{Avoiding Death by Vacuum}

\author{A. Barroso$^1$, P.M. Ferreira$^{1,2}$, I. Ivanov$^{3,4}$, R. Santos$^{1,2}$ and Jo\~{a}o P.~Silva$^{2,5}$}

\address{$^1$ Centro de F\'{\i}sica Te\'{o}rica e Computacional, Faculdade de Ci\^{e}ncias,
Universidade de Lisboa, Av.\ Prof.\ Gama Pinto 2, 1649-003 Lisboa, Portugal}
\address{$^2$ Instituto Superior de Engenharia de Lisboa, 1959-007 Lisboa, Portugal}
\address{$^3$ IFPA, Universit\'{e} de Li\'{e}ge, All\'{e}e du 6 Ao\^{u}t 17, b\^{a}timent B5a, 4000 Li\'{e}ge, Belgium}
\address{$^4$ Sobolev Institute of Mathematics, Koptyug avenue 4, 630090, Novosibirsk, Russia}
\address{$^5$ Instituto Superior de Engenharia de Lisboa, 1959-007 Lisboa, Portugal}

\ead{barroso@cii.fc.ul.pt, ferreira@cii.fc.ul.pt, igor.ivanov@ulg.ac.be, rsantos@cii.fc.ul.pt, jpsilva@cftp.ist.utl.pt}

\begin{abstract}
The two-Higgs doublet model (2HDM) can have two electroweak breaking, CP-conserving, minima. The possibility
arises that the minimum which corresponds to the known elementary particle spectrum is metastable,
a possibility we call the ``panic vacuum". We present analytical bounds on the parameters of the
softly broken Peccei-Quinn 2HDM which are necessary and sufficient conditions to avoid this possibility. We also show that, for this particular model, the current LHC data already tell us that we are necessarily in the global minimum of the theory, regardless of any cosmological considerations about the lifetime of
the false vacua.
\end{abstract}

\section{Introduction}

The two Higgs doublet model~\cite{Lee:1973iz} is one of the simplest extensions of the Standard
Model (SM) of particle physics. It has the particle content of the SM {\em plus} a second Higgs doublet,
of the same hypercharge as the SM's. It has a very rich phenomenology (for a review,
see~\cite{Branco:2011iw}), including the possibility of spontaneous CP-violation, dark matter candidates
and a richer scalar spectrum, including two CP-even scalars (usually denoted $h$ and $H$), a
pseudoscalar ($A$) and a charged scalar ($H^\pm$). With the recent
discovery of the Higgs boson~\cite{:2012gk,:2012gu}, and a possibility that its observed
properties might exhibit some deviations from its expected SM behaviour, it is interesting to
look into SM extensions and compare their predictions with data. In that regard, the 2HDM has
already shown to be able to do a good job describing current LHC results~\cite{Chen:2013kt,Belanger:2012gc,Chang:2012ve,Ferreira:2011aa}.

The 2HDM also has a very rich vacuum structure: unlike
the SM, there is the possibility of occurrence of charge breaking vacua which would give mass to the
photon; and, as already mentioned, there is also the possibility of vacua which, other than the
electroweak symmetry, also boasts a spontaneous breaking of the CP symmetry.
It has been proven~\cite{Ferreira:2004yd,Barroso:2005sm}, though, that whenever a vacuum which
breaks electroweak symmetry but preserves the electromagnetic and CP symmetries exists - which we
call a ``normal" vacuum - any charge or CP breaking stationary points which might exist are
{\em necessarily} saddle points, and lie above the normal minimum. The stability of the normal
minimum against charge or CP breaking is thus guaranteed. There is however another possibility:
the 2HDM potential, under certain circumstances, can have {\em two} normal minima, which
coexist with one another~\cite{Barroso:2007rr,Ivanov:2006yq,Ivanov:2007de}. In one of those
minima, the two Higgs doublets, $\Phi_1$ and $\Phi_2$, would have vacuum expectation values (vevs)
$v_1$ and $v_2$ such that $v_1^2 + v_2^2\, =\, 246$ GeV$^2$, and all elementary particles
would have the masses we know - this would be ``our" minimum. But in the second minimum - deeper
or higher than ours - the fields would have vevs $\{v^\prime_1\,,\,v^\prime_2\}$, with
${v^\prime}^2_1\,+\,{v^\prime}^2_2\,\neq 246$ GeV$^2$ - and the elementary particles might have masses
much smaller, or larger, than what is observed.

The possibility then arises that ``our" vacuum is {\em not} the global minimum of the theory - and
the universe would therefore be in a metastable state, with the possibility of tunneling to the
deeper vacuum. We call this situation the ``panic vacuum". It is therefore interesting to ascertain:
under what conditions such vacua might occur; whether the current experimental data can tell us something
about the nature of the vacuum in the 2HDM; what regions of parameter space are free of these panic
vacua. In~\cite{Barroso:2012mj} we presented the conditions that the parameters of the potential need to obey
so one can avoid the presence of a panic vacuum in the softly broken Peccei-Quinn~\cite{Peccei:1977hh}
version of the 2HDM. In this talk, we will briefly review those theoretical bounds, as well as the
phenomenological  analysis that was performed to compare the model's predictions with the current LHC data.

\section{Panic vacuum bounds in the softly broken Peccei-Quinn 2HDM}

The most general 2HDM scalar potential has, after all possible simplifications, 11 independent
real parameters. But it is plagued by tree-level flavour-changing neutral currents (FCNC) on the
Higgs Yukawa interactions, which
are extremely constrained by experimental observations. To evade this problem - and increase the
predictive power of the theory, by reducing the number of free parameters - one usually imposes
a symmetry on the Lagrangian. One such example is the discrete $Z_2$ symmetry, $\Phi_1\rightarrow \Phi_1$ and
$\Phi_2\rightarrow -\Phi_2$, first proposed by Glashow, Weinberg and
Paschos~\cite{Glashow:1976nt,Paschos:1976ay}. Another possibility is to impose a global, continuous
$U(1)$ symmetry, $\Phi_1\rightarrow \Phi_1$ and $\Phi_2\rightarrow \,e^{i\alpha}\Phi_2$, for any real
value of $\alpha$. This symmetry also liquidates any tree-level FCNC but, if broken by the vacuum (which
occurs if both scalar fields develop a vev), leads to a massless pseudoscalar, an axion. To prevent
that one adds a real soft breaking term to the potential, thus obtaining an acceptable scalar spectrum.

The softly broken Peccei-Quinn potential is then written as
\begin{eqnarray}
V
&=&
m_{11}^2 \Phi_1^\dagger \Phi_1 + m_{22}^2 \Phi_2^\dagger \Phi_2
- m_{12}^2 \left[  \Phi_1^\dagger \Phi_2 + \Phi_2^\dagger \Phi_1 \right]
\nonumber\\[6pt]
& &
+ \frac{1}{2} \lambda_1 (\Phi_1^\dagger\Phi_1)^2
+ \frac{1}{2} \lambda_2 (\Phi_2^\dagger\Phi_2)^2 +
\lambda_3 (\Phi_1^\dagger\Phi_1) (\Phi_2^\dagger\Phi_2)
+ \lambda_4 (\Phi_1^\dagger\Phi_2) (\Phi_2^\dagger\Phi_1),
\label{VH1}
\end{eqnarray}
where all the parameters are real and the soft breaking term is $m_{12}^2$. It can be
seen~\cite{Barroso:2007rr,Ivanov:2006yq,Ivanov:2007de} that this model, due precisely
to the soft breaking term, can have two normal minima. If the the potential has a depth
equal to $V_N$ in the minimum with vevs $\{v_1\,,\,v_2\}$, and a depth $V_{N^\prime}$ in
the minimum with vevs $\{v^\prime_1\,,\,v^\prime_2\}$, it is possible to show that
\begin{eqnarray}
V_{N^\prime} - V_{N} &=& \frac{1}{4}\,\left[\left(\frac{m^2_{H^\pm}}{v^2}\right)_{N}
- \left(\frac{m^2_{H^\pm}}{v^2}\right)_{N^\prime}\right]\,
(v_1 v^\prime_2 - v_2 v^\prime_1)^2 \nonumber \\
 &=& \frac{m^2_{12}}{4 v_1 v_2}\,
\left(1 - \frac{v_1 v_2 }{v^\prime_1 v^\prime_2}\right)\,
(v_1 v^\prime_2 - v_2 v^\prime_1)^2\,
\end{eqnarray}
with both the charged Higgs mass $m^2_{H^\pm}$ and the sum of the squared vevs $v^2$ being
computed at each of the minima. It is not therefore obvious which is the deepest minimum.

In refs.~\cite{Ivanov:2006yq,Ivanov:2007de,ivanovPRE} generic conditions for the existence
of two minima in the 2HDM were established, being put into a simpler form in~\cite{Barroso:2012mj}.
Thus it is possible to show that the Peccei-Quinn potential can have two normal minima
if the following conditions are met:
\begin{eqnarray}
m_{11}^2 + k^2\, m_{22}^2
&<& 0,
\label{M_0}
\\
\sqrt[3]{x^2} + \sqrt[3]{y^2}
&\leq&  1,
\label{astroid}
\end{eqnarray}
where the variables $x$ and $y$ are given by
\begin{eqnarray}
x
&=&
\frac{4\ k\ m_{12}^2}{
m_{11}^2 + k^2\, m_{22}^2}\,
\frac{\sqrt{\lambda_1 \lambda_2}}{
\lambda_{34} - \sqrt{\lambda_1 \lambda_2}},
\nonumber\\
y
&=&
\frac{m_{11}^2 - k^2\, m_{22}^2}{
m_{11}^2 + k^2\, m_{22}^2}\,
\frac{\sqrt{\lambda_1 \lambda_2} + \lambda_{34}}{
\sqrt{\lambda_1 \lambda_2} - \lambda_{34}},
\label{xANDy}
\end{eqnarray}
and we have defined
\begin{equation}
\lambda_{34} = \lambda_3 + \lambda_4
\hspace{5mm}
\textrm{and}
\hspace{5mm}
k = \sqrt[4]{\frac{\lambda_1}{\lambda_2}}.
\end{equation}
The curve $\sqrt[3]{x^2} + \sqrt[3]{y^2} =  1$, which delimits the
region of parameter space where two minima can occur, is known as an astroid.
To verify whether the $\{v_1\,,\,v_2\}$ minimum is the global one, one simply has to
verify if the following condition is met: one computes, at that minimum, the value of
the following quantity, which we dub the ``discriminant",
\begin{equation}
D \,=\,
\left( m_{11}^2 - k^2 m_{22}^2 \right)
(\tan{\beta} - k).
\label{D}
\end{equation}
Our minimum is the global one if, and only if, $D \,>\,0$. This condition can be verified
regardless of eqs.~\ref{M_0},~\ref{astroid}. Meaning, we do not even need to know whether the
potential has two minima, this extremely simple condition is all we need to check to ascertain
the nature of the 2HDM minimum.

\section{LHC data and panic vacua}

In order to verify how often the possibility of two minima arises in the Peccei-Quinn
potential, we performed a thorough scan of the model's parameter space. To wit,
we have considered $m_h = 125$ GeV, $125 < m_H < 900$ GeV, $90 < m_A, m_{H^\pm} < 900$ GeV,
$-\pi/2 < \alpha < \pi/2$, $1 < \tan\beta < 40$ and $|m^2_{12}| < 900$ GeV$^2$.
So that the potential is bounded from below, the quartic couplings of the potential must
obey
\begin{eqnarray}
\lambda_1 > 0 & , &  \lambda_2 > 0 \; ,\nonumber \\
\lambda_3 > -\sqrt{\lambda_1 \lambda_2} & , &
\lambda_3 + \lambda_4 - |\lambda_5| > -\sqrt{\lambda_1 \lambda_2} \;.
\label{eq:bfb}
\end{eqnarray}
We further demanded that the quartic couplings be such that the model satisfies perturbative unitarity~\cite{Kanemura:1993hm,Akeroyd:2000wc} and the electroweak
precision constraints stemming from the S, T and U parameters~\cite{Peskin:1991sw,lepewwg,gfitter1,gfitter2}.
All of these bounds apply to the theory's scalar sector. But the $U(1)$ symmetry we imposed,
even though softly broken, needs to be a symmetry of the full Lagrangian, otherwise we would
end up with a non-renormalizable model. There are several ways to extend this symmetry from
the scalar to the fermion sector. Here we will consider the two most studied: Model I, where
the fermion fields transform under the global $U(1)$ in such a way that only $\Phi_2$ couples
to the fermions; and Model II, where $\Phi_2$ couples only to up-type quarks, and $\Phi_1$ to
down-type quarks and charged leptons. Both of these models possess very different phenomenologies.
Also, data stemming from B-physics measurements put severe constraints on each of them. We have
taken those bounds into account, incorporating them in our simulations~\cite{bphys1,bphys2}.

\begin{figure}
\begin{center}
\includegraphics[width=8cm]{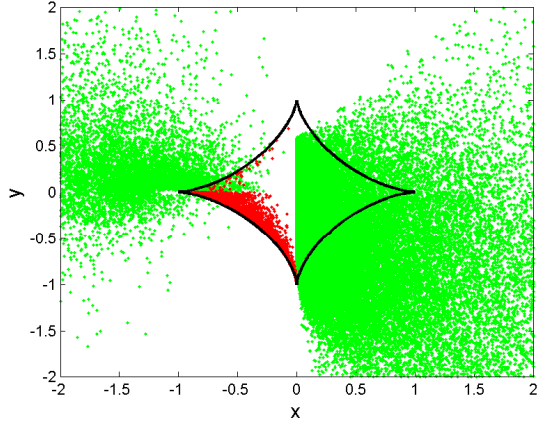}
\end{center}
\caption{\label{fig:astroid}
Distribution of generated points in the $x$ and $y$ variables. The dark curve is the
astroid described by eq.~\ref{astroid}. Points inside that curve can have two minima.
The red/dark grey points shown here correspond to panic vacua. The green/light grey points
correspond to either the non-existence of two minima, or the case where our vacuum
is the global one.
}
\end{figure}

The relevance of the conditions~\ref{M_0}-~\ref{astroid} can be appreciated in
figure~\ref{fig:astroid}. For the totality of the parameter space we scanned, we plot,
in the $\{x\,,\,y\}$ plane of the variables defined in eq.~\ref{xANDy}, the points
which correspond to the panic vacua situation (in red/dark grey) and those for which
our minimum is ``safe" - whether because there is a single minimum, or because the
second minimum is above ours. The point we wish to stress is that the occurrence of
panic vacua is not a curiosity of the model: for a blind scan of the parameter space,
obeying all reasonable constraints, a very large quantity of points with panic vacua
appears.

\begin{figure}
\begin{center}
\includegraphics[width=8cm]{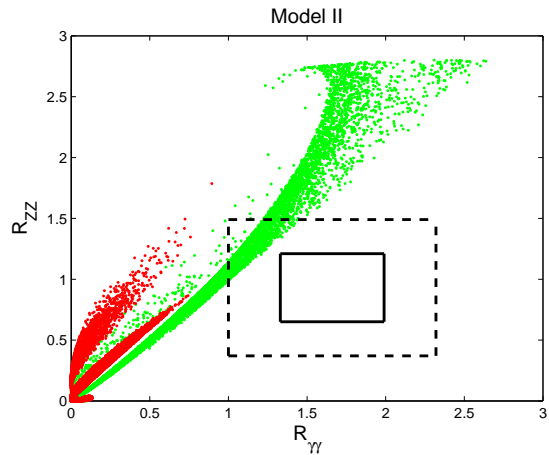}
\end{center}
\caption{\label{fig2}
$R_{ZZ}$ versus $R_{\gamma\gamma}$ for Model II. We show rough 1-$\sigma$ (solid line) 
and 2-$\sigma$ (dashed line) current bounds obtained from LHC data. Green/light grey points are the
totality of our simulation; red/dark grey points correspond to panic vacua. 
}
\end{figure}
We then use the generated points to compute $R_f$, defined as the number of events
predicted in the 2HDM for the process $pp \rightarrow h \rightarrow f$, for some final
state $f$, divided by the prediction obtained in the SM for the same final state.
Current LHC bounds at $1 \sigma$, which we took from~\cite{average}, are
$R_{ZZ} = 0.93 \pm 0.28$, $R_{\gamma\gamma} = 1.66 \pm 0.33$. We
sum over all production mechanisms, such as gluon-gluon fusion, vector boson
fusion (VBF), associated Higgs production (with a gauge boson or a $t\bar{t}$
pair) or $b\bar{b}$ fusion. In fig.~\ref{fig2} we show, for Model II, the observables
$R_{ZZ}$ versus $R_{\gamma\gamma}$, where the points corresponding to panic vacua
are shown in red/dark grey. We see that Model II is capable, at the 2-$\sigma$ level,
to obey current experimental bounds. However, even at that level, all points corresponding
to panic vacua are disfavoured by current bounds - remarkably, even though the LHC
hasn't found (yet) any evidence for more than one scalar, the current data is already
capable of telling us a lot about the vacuum structure of the 2HDM. We see that this
\begin{figure}
\begin{center}
\includegraphics[width=8cm]{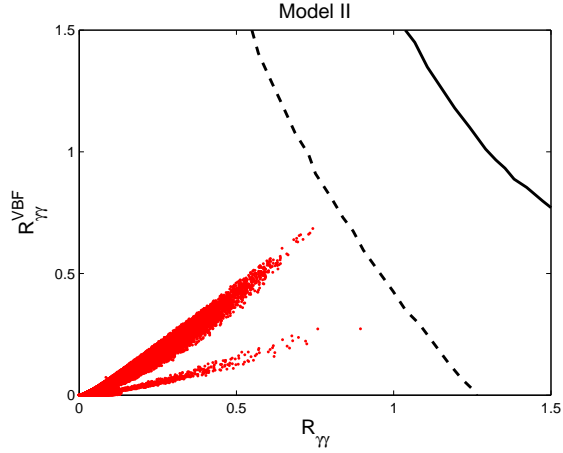}
\end{center}
\caption{\label{fig3}
$R^{VBF}_{\gamma\gamma}$ versus $R_{\gamma\gamma}$ for Model II. We show rough 1-$\sigma$ (solid line)
and 2-$\sigma$ (dashed line) current bounds obtained from LHC data. The red/dark grey points correspond to panic vacua.
}
\end{figure}
trend is also verified in other variables, such as those plotted in fig.~\ref{fig3}. There
we plot the Higgs $\gamma\gamma$ event rate (relative to the SM's), produced only via the VBF
mechanism versus its total production rate. We only show those points corresponding to
the panic vacuum solutions, and see that, even at the 2-$\sigma$ level, they are strongly
disfavoured.

\begin{figure}
\begin{center}
\includegraphics[width=8cm]{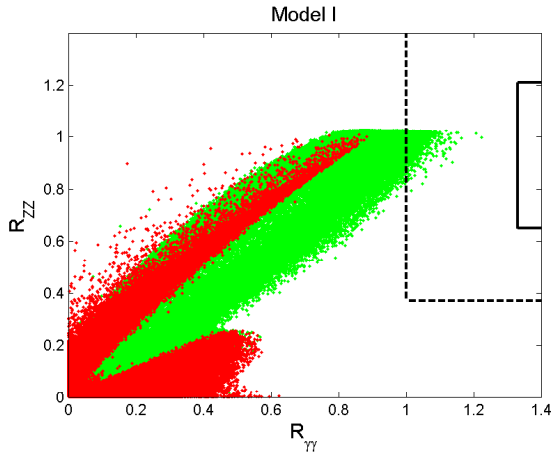}
\end{center}
\caption{\label{fig4}
$R_{ZZ}$ versus $R_{\gamma\gamma}$ for Model I. We show rough 1-$\sigma$ (solid line)
and 2-$\sigma$ (dashed line) current bounds obtained from LHC data. Green/light grey points are the
totality of our simulation; red/dark grey points correspond to panic vacua.
}
\end{figure}
The same, however, cannot be said to occur in Model I. Even though, as we see in
fig.~\ref{fig4}, where we once again plot $R_{ZZ}$ versus $R_{\gamma\gamma}$, all
panic points (the red/dark grey ones) lie outside even the 2-$\sigma$ band, in
the $R^{VBF}_{\gamma\gamma}$-$R_{\gamma\gamma}$ plane that would not occur: the
2-$\sigma$ band would include some panic points, even though that would not occur at
the 1-$\sigma$ level. Of course, the fact that all panic points are excluded due to
the results of fig.~\ref{fig4} would mean that for Model I, as well, the existence
of panic vacua is strongly disfavoured by current LHC data, albeit perhaps less so than
in Model II.

\section{Conclusions}

The 2HDM can have metastable neutral vacua, and this raises the possibility that
the minimum we are currently inhabiting could not be the global minimum of the
model. We have presented the extremely simple conditions one has to impose on the
parameters of the potential to prevent this situation. They are, we believe, to be
taken as seriously as the bounded-from-below conditions one usually imposes on the
potential, or those enforcing perturbative unitarity. These bounds can also be extended to other
versions of the 2HDM, and that work is underway~\cite{z2}.

The argument can be made that the existence of a metastable vacuum is not a problem
{\em per se}, if the tunneling time from the false vacuum to the true one is larger
than the current age of the universe. We have performed a quick estimate of the lifetimes
of our panic vacua, and concluded that most of them would have lifetimes inferior to
the age of the universe, therefore they would correspond to truly undesirable regions
of parameter space, in disagreement with observed phenomenology.

But the most interesting aspect of our analysis, we believe, is the fact that, regardless
of any calculation of vacuum lifetimes - and it must be stressed that that calculation
is rife with approximations and considerable assumptions - the current data stemming from
the LHC already permit us to conclude a lot about the nature of our vacuum, if Nature is
indeed described by the 2HDM. Namely, the present LHC bounds on several observables already
tells us to strongly disfavour the possibility that the vacuum we are in is {\em not}
the global minimum of the theory. But that does not diminish the validity of the panic
vacuum bounds we presented here, nor their interest: for we see that panic vacua
occur for perfectly ordinary values of the 2HDM parameters, which predict values for
observables which are not {\em a priori} absurd. And in any case, we see that the 
discriminant of eq.~\ref{D}, by itself, allows us to obtain information, via only particle
physics experiments, about a very interesting cosmological subject: the nature of the 
universe's vacuum.

\ack
The works of A.B., P.M.F. and R.S. are supported in part by the Portuguese
\textit{Funda\c{c}\~{a}o para a Ci\^{e}ncia e a Tecnologia} (FCT)
under contract PTDC/FIS/117951/2010, by FP7 Reintegration Grant, number PERG08-GA-2010-277025,
and by PEst-OE/FIS/UI0618/2011.
The work of J.P.S. is funded by FCT through the projects
CERN/FP/109305/2009 and  U777-Plurianual,
and by the EU RTN project Marie Curie: PITN-GA-2009-237920.
I.P.I. is thankful to CFTC, University of Lisbon, for their hospitality.
His work is supported by grants RFBR 11-02-00242-a,
RF President grant for scientific schools NSc-3802.2012.2, and the
Program of Department of Physics SC RAS and SB RAS "Studies of Higgs boson and exotic particles at LHC".

\end{document}